\documentclass{aa}  

\usepackage{graphicx}
\usepackage{txfonts}
\usepackage[]{hyperref}
\usepackage{xcolor}

\newcommand{\modif}[1]{{#1}}
\newcommand{\Rsun}{\mathrm{R}_\odot}
\begin{document}

   \title{On the feasibility of inverting the rotation of the solar core with mixed $f/g$ modes}


   \author{A. Leclerc
          \inst{1} \thanks{These authors contributed equally to this work}\fnmsep\thanks{Corresponding author: armand.leclerc@ista.ac.at}
          \and
          A. Le Saux\inst{2 ~ \star}
          \and 
          J. M. J. Ong\inst{3, 4~ \star}
           \and 
          R. A. García\inst{2}
          }

   \institute{ Institute of Science and Technology Austria, Am Campus 1, Klosterneuburg, 3400, Austria
   \and
   Université Paris-Saclay, Université Paris Cité, CEA, CNRS, AIM, Gif-sur-Yvette, F-91191, France.
   \and
   Institute for Astronomy, University of Hawai’i, 2680 Woodlawn Drive, Honolulu, HI 96822, USA
   \and
   Sydney Institute for Astronomy, School of Physics, University of Sydney, A28 Physics Road, Sydney, NSW 2006, Australia
   }
   \date{Received Month 99, 2026; accepted Month 99, 202x}

 
  \abstract
   {Thanks to helioseismology, the rotation profile of the Sun has been measured with great precision down to 20\% of its total radius. This rotation profile is used as a calibration to infer the rotation of other stars as well as a test of angular momentum transport theory in stellar interiors. However, the {deepest 20\% of the }layers remain out of reach of current observations, preventing astronomers to discriminate between currently competing angular momentum transport mechanisms. }
   {The main obstacle is that no global oscillations modes sensitive to rotation ($\ell>0$) reaching the solar core have been detected yet, as nonradial $p$ modes cannot reach it and $g$ modes are evanescent at the surface and still elude detection. In this work, we propose and examine a new method to constrain the rotation of the core of the Sun, which does not require direct observation of solar $g$ modes.}
   {It is based on a recent prediction that $g$ modes in the radiative interior couple with $f$ modes in the outer parts of the star. These mixed $f$/$g$ are at the same time sensitive to the rotation of the core and able to reach the surface. These modes can be used together to build average inversion kernels and perform an inversion of the rotation of the solar core.}
   {We find that the oscillations' spectrum of the Sun should present 6 mixed $f$/$g$ modes that can be used to measure the rotation rate of the Sun at $r = 0.07$ and $0.2 \Rsun$. We estimate that the uncertainty on the measurements should be small enough to distinguish between competing scenarios of angular momentum transport in the Sun.}
   {}

   \keywords{Sun --
                Rotation --
                Helioseismology
               }

   \maketitle
%

\section{Introduction}
\label{sec:introduction}
The global oscillation modes of the Sun have been studied in great detail since the first detection of the five-minute oscillations by \citet{Leighton1962}. In the following decades, helioseismology has allowed to constrain with high precision the internal structure and dynamics of our nearby star \citep[see e.g. the review by][]{JCD2021} using its global acoustic modes. The modes with angular degree $\ell$ > 0 are degenerate in azimuthal order $m$. The rotation of the star lifts this degeneracy by inducing a shift in frequency of an oscillation mode, which depends on the order $m$ of the mode. By measuring this shift, called rotational splitting, helioseismologists have been able to infer the rotation profile of the Sun in great detail \citep[e.g.][]{Thompson2003, Howe2009}. It has revealed the latitudinal differential rotation of the convective envelope and the solid-body rotation of the radiative interior down to $r \sim 0.2\Rsun$ \citep{chaplin1999rotation,Couvidat2003, Garcia2011, Korzennik2024}. However, these modes, also called $p$ modes, do not allow us to probe the rotation in the innermost layers of the Sun as they do not propagate so deep. 
In this region are trapped gravity modes, or $g$ modes, which are thus very sensitive to the dynamics of the core \citep{mathur2008sensitivity}. Unfortunately, $g$ modes are evanescent in the convective envelope, so their amplitude at the surface of the Sun is extremely difficult to detect, and inversion techniques of the core are very uncertain \citep{eff2008analysis}.
Despite this challenge, some studies have reported the detection of solar $g$ modes. Almost twenty years ago, \citet{Garcia2007} detected a signal that was coherent with the period spacing for $g$ modes. Although being robust to subsequent analysis, the physical origin of the signal was questioned by \citet{Appourchaux2010}. More recently, \citet{Fossat2017}, reported the detection of $g$ modes signature in $p$ modes oscillations. However, when trying to reproduce the result, \citet{Schunker2018, Appourchaux2019} did not obtain the same signature in the frequencies of the $p$ modes, which cast doubts on their detection. Interestingly, despite using very different methods, both \citet{Garcia2007} and \citet{Fossat2017} predicted that the core of the Sun rotates faster than the rest of the radiative interior. Nevertheless, there is no clear detection of individual solar $g$ modes yet, and measuring the rotation rate of the core remains one of the most important challenges in solar physics.

In order to explain the solid-body rotation in the radiative interior, a very efficient redistribution of angular momentum is needed. Two mechanisms have been proposed to explain this rotation profile: internal gravity waves \citep{Charbonnel2005} and internal magnetic field \cite{fuller2019,eggenberger2019rotation}.
For the latter, two scenarios are proposed to explain the solid-body rotation down to $r\sim 0.2\Rsun$, both based on the Tayler-Spruit instability, but their prediction differs for the rotation rate of the solar core. {While} \cite{fuller2019} proposes a very efficient angular momentum transport and thus expects solid-body rotation down to the centre of the Sun, \cite{eggenberger2019rotation} predicts instead that the core spins approximately three times faster than the rest of the radiative zone. This contradiction between these scenarios prevents the development of a unified theory of angular momentum transport in stellar interiors, and thus limits our predictions for other stars. The only way to solve this tension is by measuring the rotation rate of the core of the Sun.

{Avoided crossings between low frequency $f$ modes and low order $g$ modes have been studied in the past \cite{christensen1980adiabatic}, with a recent renewed interest when} \cite{LeSaux2025} predicted a mixed $f$/$g$ mode in the spectrum of the Sun, which has amplitude in its core as well as at its surface. This new kind of oscillation mode should thus be detectable as well as sensitive to the rotation rate of the core. Located at higher frequency, where the noise from convection motions is lower, it could be an easier target for observations than pure $g$ modes, while offering similar sensitivity to the dynamics of the core. The study of \cite{LeSaux2025} predicts such a mixed mode at $\ell = 4$ and a frequency close to 265 $\mu$Hz. However, one should expects a potential mixed $f/g$ mode for each harmonic degree $\ell$. In the present work, we convey a systematic search of these mixed $f$/$g$ modes. In Sect. \ref{sec:mixed} we assess how many of these mixed modes can be expected for the Sun, and then estimate how they can be used conjointly to {constrain} the rotation profile on the solar core in Sect. \ref{sec:avg-ker}. Finally, in the last section we determine which one of these mixed modes offers the most promising observational target, before concluding.\\

\section{Properties of all solar mixed f/g modes}
\label{sec:mixed}
A mixed $f$/$g$ mode results from the coupling of a fundamental $f$ mode in the outer part of the Sun, and a gravity $g$ mode trapped in the radiative interior.
Recently, \cite{LeSaux2025} predicted such a mixed $f$/$g$ modes in the seismic spectrum of the Sun using wave topology and fully compressible hydrodynamical simulations. This coupling occurs as the $f$ modes branch penetrates in the frequency range of the $g$ modes band at low angular degree. The authors focused on the $\ell = 4$ mode, but as one can notice by looking at a {numerically computed} solar spectrum, the $f$ and $g$ modes' frequencies overlap for all angular degrees $\ell \lesssim $ 20. Thus, other mixed $f$/$g$ modes should exist at different angular degrees.
In this section, we determine all the mixed $f$/$g$ modes that can be expected for the Sun. Then for each one, we estimate if the coupling is sufficiently strong such that the sensitivity to the core rotation rate is high enough in order to get precise constraints of the rotation rate of the solar core.
The 1D model we use for this study is the calibrated solar model introduced in \citet{LeSaux2025}. This model was build using the Modules for Experiments in Stellar Astrophysics (MESA) stellar evolution code \citep{Paxton2011, Paxton2013, Paxton2015, Paxton2018, Paxton2019,Jermyn2023}. The eigenfunctions and eigenfrequencies of this 1D model are computed using the oscillations code GYRE \citep{Townsend2013, Townsend2018}.

\begin{figure*}
    \centering
    \includegraphics[width=\columnwidth]{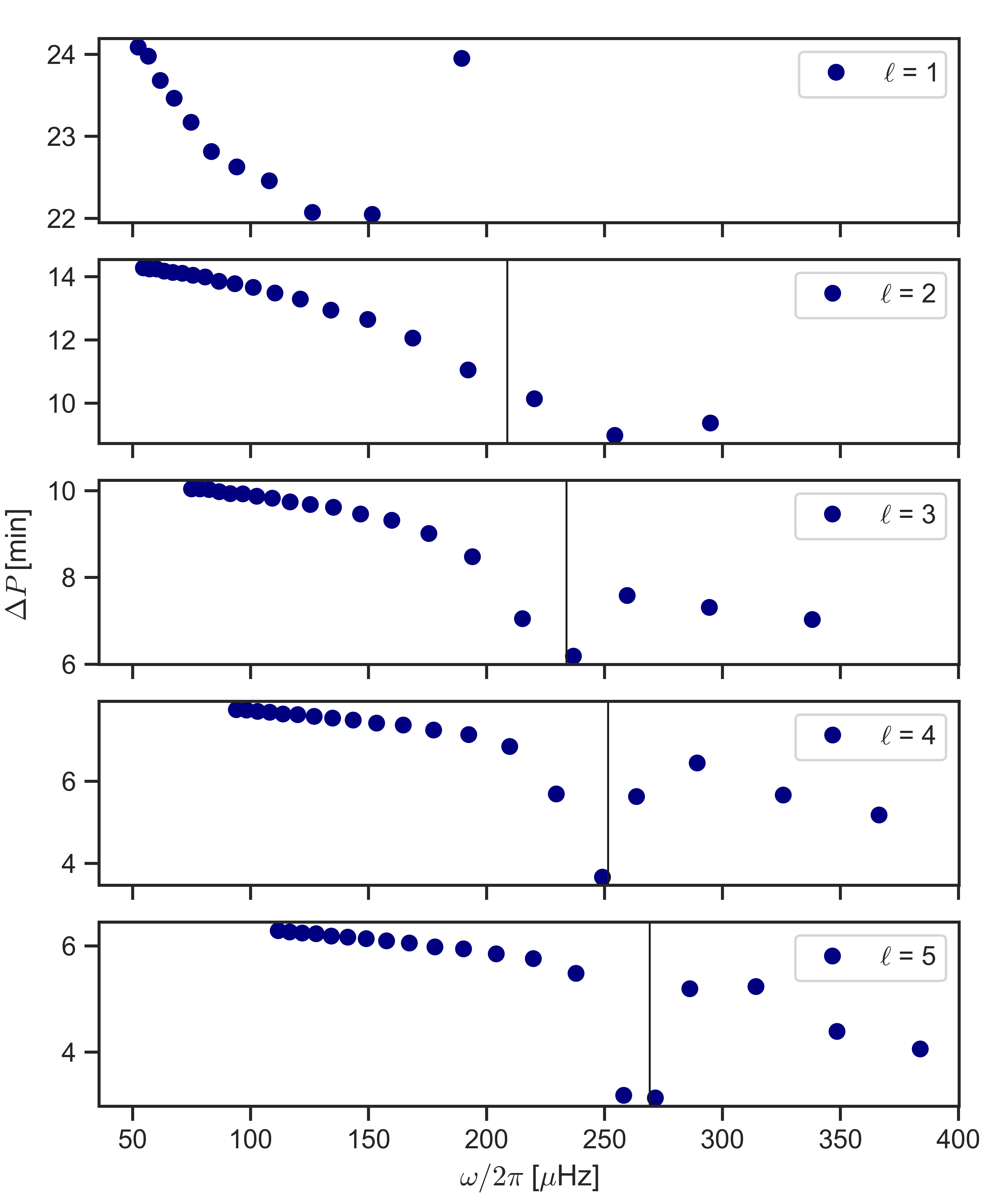}
    \includegraphics[width=\columnwidth]{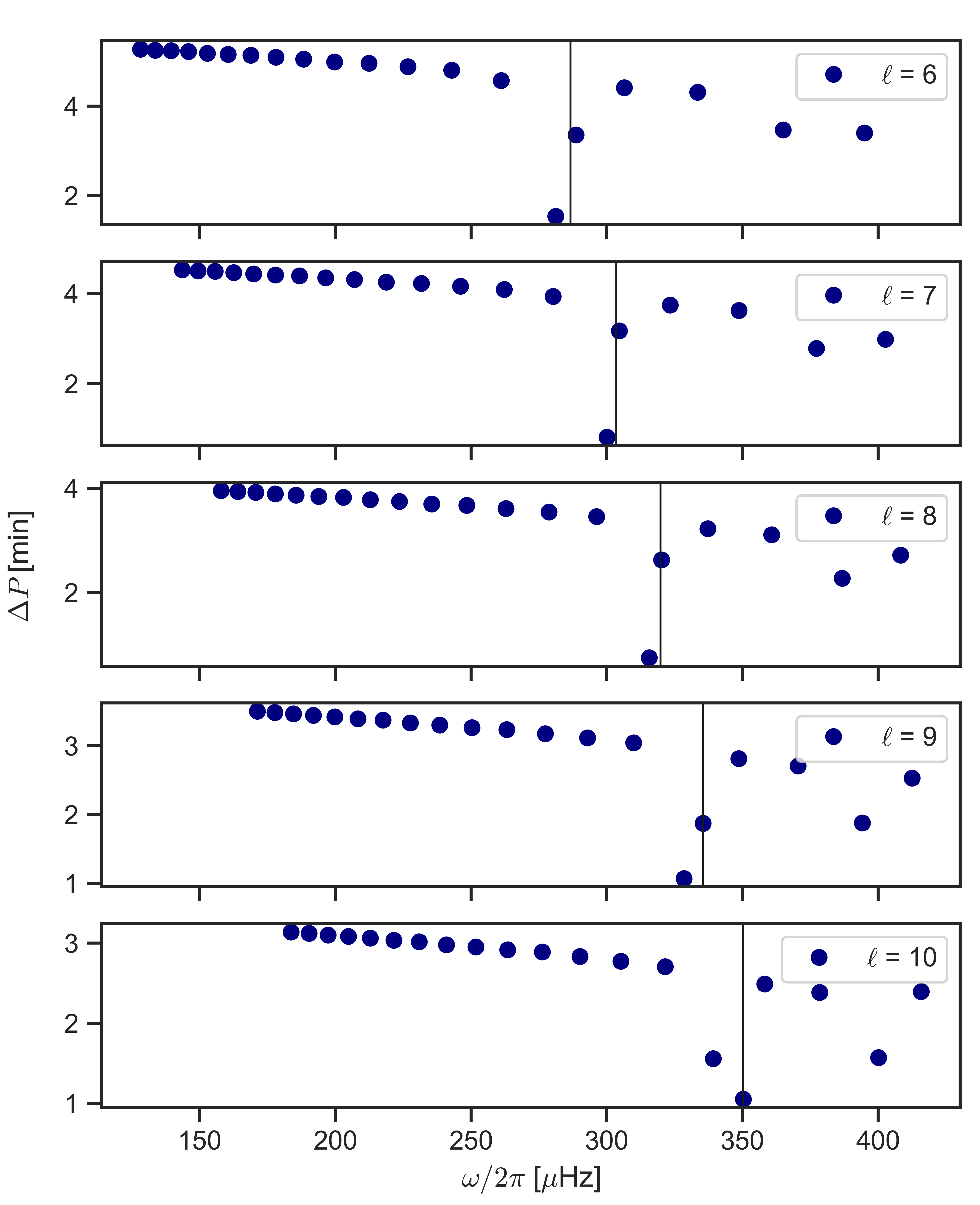}
    \caption{Consecutive period differences $\Delta P$ as a function of frequency $\nu$ for solar oscillations modes with angular degrees between $\ell$ = 1 and 10. {The frequency of the $\varphi$ mode is identified by the vertical black lines}. 
    }
    \label{fig:delta_pi}
\end{figure*}

A commonly used diagnostic to determine if a $g$ mode can couple to $p$ modes in evolved stars is to look at the period spacing pattern. Indeed, it is known that $g$ modes of consecutive radial order are equally spaced in period \citep{Tassoul1980}. Therefore, if a coupling is susceptible to occur, an extra mode, which would be a $p$ mode in the case of evolved stars and a $f$ mode in our case, will create a dip in this constant period spacing pattern. Figure \ref{fig:delta_pi} presents this diagnostic for modes of angular degrees $\ell$ between 1 and 10 and radial orders $n$ between 0 and -20. The period spacing is defined as $\Delta P = P_{n+1} - P_{n}$, with $P_n = 2\pi/\omega_{\ell,n}$ the period of the mode with frequency $\omega_{\ell,n}$. All the blue dots correspond to $g$ modes, with increasing radial order towards lower frequencies, except for the dot identified with the vertical black line for degrees between 2 and 10, which is the $f$ mode of corresponding angular degree. {These vertical lines correspond to pure $f$ mode frequencies, i.e. when neglecting the coupling with $g$ modes, which are computed using the decomposition procedure for mixed modes described in \citet{Ong2020}}. {We describe the theoretical construction and numerical procedure for this decomposition more completely in \autoref{app:decomp}}.  As in that work, the wave operator generating mixed $p$/$g$ modes can be decomposed into two parts: one, the $\pi$ mode subsystem, generates pure $p$ modes, while the other, the $\gamma$ mode subsystem, generates pure $g$ modes. As shown by \citet{Aizenman1977}, the $\pi$ mode subsystem also contains a single $f$-like mode --- the $\varphi$ mode --- despite this mode corresponding to propagation of a gravity wave; this is because the $\pi$ mode isolation conditions of both \citet{Aizenman1977}, and \citet{Ong2020}, are constructed only to suppress gravity-wave propagation interior to the convective envelope.
In particular, we have used the same procedure to derive an $f$ mode frequency from the $\varphi$ mode eigenfunction as that to derive pure $p$ mode frequencies from $\pi$ modes. We also use this procedure to compute coupling matrices --- which are overlap integrals of the wave operator between different pure $p$, $f$, and $g$ modes that couple this $\varphi$ mode to $\gamma$ modes to yield mixed $f$/$g$ modes --- using the same procedure as described in \citet{Ong2020}. These yield a characteristic resonance width $\Gamma$ for each $f$-mode, which is defined as
\begin{equation}
    \Gamma = {1 \over \pi \Delta\Pi_\ell} \left(\alpha \over 8\pi\nu^2\right)^2,
    \label{eq:resonance_width}
\end{equation}
where $\alpha$ is the coupling strength between the $f$ and the $g$ modes \citep{Ong2023}. 

\begin{figure}
    \centering
    \includegraphics[width=\columnwidth]{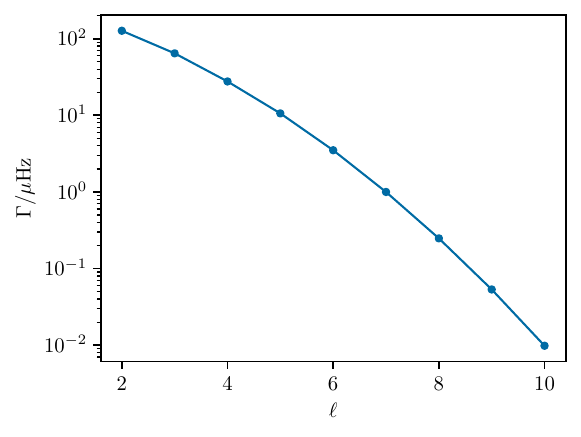}
    \caption{Resonance widths of solar mixed $f$/$g$ modes. For $\ell \leq 8$, the width is larger than frequency splittings and surface-effects corrections, consistently with these modes indeed being coupled in the Sun.}
    \label{fig:resonance-widths}
\end{figure}

It appears clearly that these $f$ modes create a dip in the period spacing pattern, suggesting the existence of mixed $f$/$g$ modes for all these degrees. For $\ell$ = 1, there is not any $f$ mode as this would be an unphysical case where the centre of mass of the star is displaced \citep{unno1979}. 
For $\ell$ = 2, the dip does not appear clearly. We suggest that it is because this dip is relatively wide in terms of frequency. Indeed, from Fig. \ref{fig:resonance-widths}, we can see that the resonance width of the dip, as defined in Eq. \eqref{eq:resonance_width}, increases towards lower angular degrees. This causes the dip to be the widest for $\ell =2$, and also the least prominent. 
One might notice that for $g$ modes of the lowest degrees, the period spacing does not appear to be constant with frequency. This is because this constant behaviour is predicted from asymptotic theory that is only valid for large values of $n$. 
By looking at Fig. \ref{fig:delta_pi}, we can clearly see dips in the period spacing pattern at frequencies close to the one of the $f$ mode for $\ell$ between 3 and 10. We have also calculated this period spacing pattern for modes with $\ell$ between 11 and 20, and find that modes with angular degrees larger than 10 cannot be used for rotational inversion as we will see in the following. Thus, we only keep modes with $\ell$ < 11. 

Next, we want to estimate the sensitivity of each mode to the solar core rotation rate $\Omega_\mathrm{c}$. 
The Sun is a slowly rotating star and, as such, the splitting of oscillations modes that rotation induces can be described as a linear perturbation to the nonrotating mode frequency, as
\begin{equation}
    \omega_{\ell,m,n} = \omega_{\ell,n} + \Delta \omega_{\ell,m,n} = \omega_{\ell,n} + m \Delta \omega_{\ell,n} + \mathcal{O}(m^2),
\end{equation}
with $\omega_{\ell,m,n}$ the frequency of the mode identified with its angular degree $\ell$, azimuthal $m$ and radial $n$ orders, which can be expressed as a function of $m$ for each $\ell, n$. This splitting can be related to the integral over the internal solar rotation rate $\Omega(r,\theta)$ weighted by the rotational kernel \citep{Howe2009} as
\begin{equation}
        \Delta\omega_{\ell,m,n} = \int K_{\ell,m,n}(r, \theta) \Omega(r,\theta)\mathrm{d}r\mathrm{d}\theta,
\end{equation}
with $r$ and $\theta$ the radial and latitudinal coordinates. 
{The largest splitting occurs for sectoral modes (i.e $\ell = m$ modes), which are mostly sensitive to the radial structure of the rotation profile }
\begin{equation}
    \Delta\omega_{\ell,\ell, n} \sim \int K_{\ell,\ell,n}(r) \Omega(r) \ \mathrm dr,
\end{equation}
where $\Omega(r)$ is a horizontal average of $\Omega(r,\theta)${, weighted by $(\sin\theta)^\ell$}.
As introduced in \citet{LeSaux2025}, {\modif{rotational} kernels allows one to define a sensitivity parameter $s$ as}
\begin{equation}
    s \equiv \frac{1}{m}\frac{\partial \omega_{\ell,m,n}}{\partial \Omega_\mathrm{c}} = {1 \over m} \int_{r<0.2 \Rsun} K_{\ell,m,n} \;\mathrm{d}r\mathrm{d}\theta,
\end{equation}
with $K_{\ell,m,n}$ the rotational kernel of the mode. Our definition of $s$ here differs from the one used by \citet{LeSaux2025} by a factor of $1/m$. The expression of $K_{\ell,m,n}$ for a given mode ($\ell$,$m$,$n$) is given for example in \citet{schou1994}. The kernels allow us to estimate how sensitive a mode is to rotation at a given radius inside the Sun. Then, we integrate its value in the core ($r < 0.2\Rsun$) to estimate the sensitivity $s$.
From an observational perspective, \modif{if there is an uncertainty $\delta\omega$ on the detection of a mode with sensitivity $s$, it will yield the core rotation rate} an uncertainty of $\Omega_c$ of $\delta\Omega_c = \delta\omega/s$.
Therefore, sensitive modes yield precise measures. As a way to select the observationally relevant mixed modes \modif{and knowing that current observational uncertainties in the BiSON dataset reach $\delta\omega\sim10$ nHz \citep{howe2023low}}, we choose to keep the modes that have $s>0.2$, such that the uncertainty on the core rotation rate will be no greater than $50$ nHz. This sensitivity threshold is much higher than the sensitivity of low degree $p$ modes, which are the ones that have higher sensitivity to the solar core and which have $s \sim 0.07$ \citep{LeSaux2025}. The values of $s$ for each mode is given in Table \ref{tab:sensitivity}.

To estimate the extent of coupling between the $f$ and $g$ modes, we compute the $g$ mode mixing fraction $\zeta$ in the same manner as described in \citet{Ong2020} for mixed $p$/$g$ modes. Mode coupling produces normal modes which are linear combinations $\xi = \sum_{i} c_{g,i}\xi_{g,i} + c_f \xi_f$ of pure $f$- and $g$ modes as eigenvectors of these coupling matrices, and $\zeta$ for each mode is computed as the squared contribution of the pure $g$ modes to that normal mode, $\zeta = \sum_i |c_{g,i}|^2$. Values of $\zeta$ close to 1 correspond to pure $g$ modes, those close to 0 to pure $f$ modes, whilst intermediate values indicate that the two modes are coupled and close to resonance. For each mode, we also compute the normalised mode inertia $E_{\rm norm}$ using GYRE. This quantity is a good proxy to estimate if a mode could be more or less easy to detect in observations. Larger values of $E_{\rm norm}$ indicate modes that are more difficult to detect \citep{Garcia2019}. These two quantities are related: if $E_f$ is the mode inertia of a pure $f$ mode, $\zeta = 1 - E_f / E_{\rm norm}$.

\begin{table}[h!]
\caption{\label{t7}Properties of the two most f-dominated mixed $f$/$g$ modes for each $\ell$ between $\ell$ = 3 and 10, computed from the solar model of \cite{LeSaux2025}.}
\centering
\begin{tabular}{lcccc}
\hline\hline
$\ell$&$\nu$ ~($\mu$Hz)&s&$E_{\rm norm}$&$\zeta$\\
\hline
3       &236     &0.45&  $1.60 \times 10^{-2}$ & 0.63\\
3       &259     &0.48&  $1.75 \times 10^{-2}$ & 0.79\\
4       &249     &0.40&  $9.42 \times 10^{-3}$ & 0.58\\
4       &263     &0.41&  $8.92 \times 10^{-3}$ & 0.65\\
5       &271     &0.17&  $3.39 \times 10^{-3}$ & 0.26\\
5       &285     &0.57&  $2.17 \times 10^{-2}$ & 0.90\\
6       &281     &0.57&  $1.53 \times 10^{-2}$ & 0.87\\
6       &289     &0.11&  $2.15 \times 10^{-3}$ & 0.18\\
7       &299      &0.59&  $1.59 \times 10^{-2}$ & 0.91\\
7       &305      &0.06& $1.48 \times 10^{-3}$ & 0.11\\
8       &315      &0.62&  $5.45 \times 10^{-2}$ & 0.98\\
8       &320      &0.01& $1.04 \times 10^{-3}$ & 0.03\\
9       &328      &0.63&  $4.27 \times 10^{-1}$ & 1.0\\
9       &335      &0.00&  $7.98 \times 10^{-4}$ & 0.0\\
10      &338      &0.63&  $3.81$ & 1.0\\
10      &350      &0.00&  $6.35 \times 10^{-4}$ & 0.0\\
\hline
\end{tabular}
\tablefoot{$s$ is the sensitivity to solar core rotation (r<0.2 R$_{\odot}$), $E_{\rm norm}$ is the mode inertia normalised at r = $R_{\rm star}$ as computed by GYRE and $\zeta$ is the mixing fraction.}
\label{tab:sensitivity}
\end{table}

The results of our inspection and characterizations of solar mixed $f$/$g$ modes are presented in Table~\ref{tab:sensitivity}. Each mixed $f$/$g$ mode is identified with its angular degree $\ell$ and frequency $\nu$. For each angular degree, there are two most $f$-dominated modes. They correspond to the $f$ mode, and the $g$ mode closest to resonance to it, that become coupled and which thus form two mixed modes: one that is dominated by the $g$ mode (the lowest frequency one in general) component and the other dominated by the $f$ mode component.
In the last column, the value of $\zeta$ tells us that modes with $\ell \geq$ 9 are not mixed. Indeed, the value of $\zeta$ is 1 for the $g$-dominated mode and 0 for the $f$-dominated one. This suggests that these modes are essentially pure $g$ and $f$ modes. 
This result is also confirmed by examining the resonance widths, shown on Fig. \ref{fig:resonance-widths} as a function of $\ell$. The value of $\Gamma$ decreases with $\ell$, and it appears that for $\ell \geq$ 9, the resonance width is smaller than the frequency splittings and surface term corrections, which also suggests that these modes are unlikely to be mixed in the Sun. This is because, for a given coupling strength, only $g$ modes within $\Gamma$ of an $f$ mode will couple significantly to it, and vice versa. The resonance width determines the characteristic frequency scales of nonlinear avoided crossings between $f$ and $g$ modes, which may invalidate the application of perturbative techniques used in describing rotation as linear multiplet splittings (as we do below), or in standard surface-term prescriptions. When the resonance width is very small, these nonlinear effects can be avoided by restricting attention to pure $f$ and $g$ modes --- but this lack of coupling also precludes observational access to the $g$ modes cavity. When it is very large, first-order perturbations do not advance mode significantly along avoided crossings, which can then be described with linear approximations, as we do.

For modes with angular degrees $\ell \in [3,7]$, the values of $\zeta$ lie between 0 and 1, indicating that the modes exhibit both $f$-like and $g$-like character. The sensitivity $s$ is larger than 0.01, meaning that they can probe the rotation of the solar core, and their inertia is relatively small with $E_{\rm norm}$ < $5 \times 10^{-2}$, compared to pure $g$ modes for which the lowest inertia is larger than 0.1. This suggests that mixed $f$/$g$ could be more easily detected than pure $g$ modes.

In the next section, we examine prospects for exploiting the probing power of the mixed $f$/$g$ modes to measure the rotation rate of the solar core. For this purpose, we only use modes with angular degrees $\ell$ between 3 and 8, as explained in this section. We keep the $\ell$ = 8 modes in our analysis as, despite being only weakly coupled ($\zeta$ close to 0 or 1) their sensitivities to the core are still significant, and the inertia remains acceptably low. In addition, these modes have the highest frequencies in this list; as the background noise from granulation and turbulence, which prevent signal detection, decreases with increasing frequency \citep{Garcia2007,Pincon2021}, it is these modes which might be more easily accessible to observations.

\section{Inverting rotation rate with averaging kernels}
\label{sec:avg-ker}

\begin{figure}
    \centering
    

    \includegraphics[trim=.25cm .25cm .25cm .15cm,clip,width=.9\columnwidth]{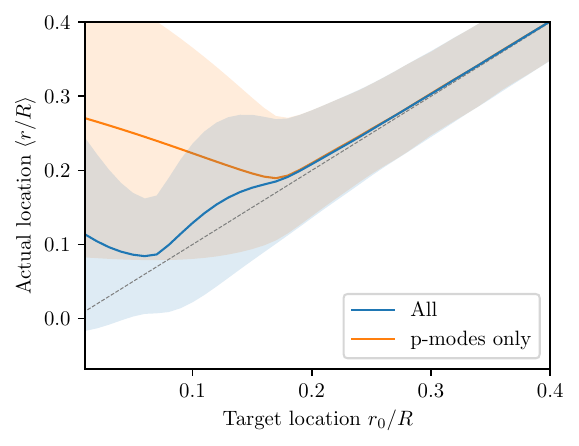}
    \includegraphics[trim=.25cm .25cm .25cm .15cm,clip,width=.9\columnwidth]{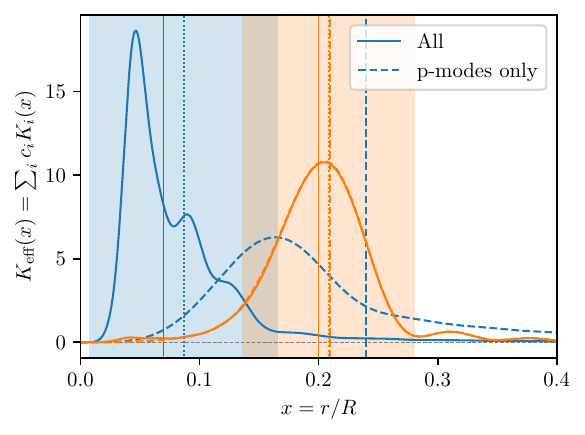}
    \includegraphics[trim=.25cm .25cm .25cm .05cm,clip,width=.9\columnwidth]{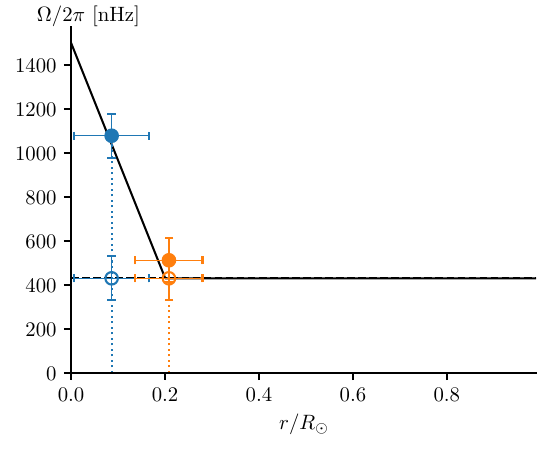}
    \caption{\footnotesize Supplementing $p$ mode rotational kernels with those of a select few $f$/$g$ modes qualitatively improves our capability to measure near-core rotation rates. \textbf{Top}: Centroids $\left<r/\Rsun\right>$ of MOLA rotational kernels (Eq.~\eqref{eq:centroid}, solid curves), plotted as a function of target locations $x_0 = r_0/\Rsun$. Shaded regions indicate the spread $\sigma$ of the rotational kernels, per Eq.~\eqref{eq:spread}. Orange curves and shaded areas show quantities calculated from kernels constructed using only $p$ modes, while blue curves and shaded areas show those calculated also including the most $f$-like modes in \autoref{tab:sensitivity}. The line of equality is shown with the grey line. \textbf{Middle}: Illustrative rotational kernels. Different colours represent different target locations ($r/\Rsun = 0.07$ and $0.2$, shown with solid vertical lines). Dashed curves show inversion kernels constructed using only p-modes, with centroids shown using dashed vertical lines. Solid curves show inversion kernels constructed with the inclusion of $f$/$g$ mixed modes, with centroids shown using dotted vertical lines. The shaded intervals represent the spread $\sigma$ of each $f$/$g$ mixed-mode kernel. \textbf{Bottom}: The two averaging kernels shown, {using $f$/$g$ modes,} can potentially discriminate between two rotation curve scenarios, depicted with the solid and dashed curves (as described in the main text). Filled circles show simulated inversions assuming the rotation curve shown with the solid curve, while open circles show those using the constant dashed curve.}

    \label{fig:avg_kernels}
\end{figure}
To estimate the rotation rate at a given radius $r_0$ inside the Sun, it has been shown that one can build average kernels \citep{schou1994}, which are a linear combination of the kernels of different modes. Specifically, we choose coefficients $c_i$ so as to specify an averaging kernel 
\begin{equation}
    K_\text{eff}(r) = \sum_{i}c_i K_{\ell_i,m_i,n_i}(r),
    \label{eq:def-avg-kernels}
\end{equation}
where $i$ indexes each individual {mode used in the average}. Such a kernel is specifically constructed to have significant amplitude, and thus probing power, at given radius $r_0$. With such a kernel, one has
\begin{equation}
    \sum_{i}c_i \Delta\omega_i = \int K_\text{eff}(r) \Omega(r)\;\mathrm{d}r \simeq \Omega(r_0),
\end{equation}
with a localisation uncertainty given by the spread of $K_\text{eff}$. This is the essence of the Optimal Local Averages (OLA) procedure (e.g. \citealt{backus1968}). Constructing a series of such average kernels allows one to map the function $\Omega(r)$, by changing the coefficients $c_i$ to probe various regions of the star (e.g. \citealt{JCD1988}).

This method is applicable when the number of modes at one's disposal is enough to be able to construct the decompositions Eq.~\eqref{eq:def-avg-kernels}, the ideal case being disposing of an infinite number of modes whose kernels $K_i$ form a complete basis. For the Sun in particular, the $p$ mode kernels {are sufficiently close to a complete basis of radial functions as to permit an accurate inversion of the profile.} 
However, because $p$ modes primarily propagate in the outer layers of a star (and increasingly so at high $\ell$), such measurements have likewise been mostly restricted to the outer layers, about $r \gtrsim 0.2 \Rsun$ \citep{mathur2008sensitivity}.

Observational access to these $f$/$g$ mixed modes may change our capability for internal rotational measurement. To illustrate this, we construct localisation kernels for the rotation rate using the method of optimally localised averages (OLA, \citealt{backus1968}). We do so in order to avoid needing to assume anything about the rotational profile a priori (as in \citealt{mathur_structure_2009}). We describe in \autoref{app:coefs} our modification of the standard OLA technique, which were required to accommodate the inclusion of these $f$/$g$ mixed modes, and specifically information about their mixing fractions $\zeta$. With these modifications in hand, we examine, in \autoref{fig:avg_kernels}, the properties of MOLA localisation kernels constructed using either only $p$ mode rotational kernels, or with the inclusion of $f$/$g$ modes. For the purposes of representing observational uncertainties in the inverted rotational profile, we use measurement errors reported in the HMI solar rotational-splittings data set \citep{korzennik_precision_2023} for $\Delta\omega/2\pi$. We restrict our attention to p-mode multiplets who splittings were reported in \cite{korzennik_precision_2023}, supplemented by the $f$-dominated mixed modes listed in \autoref{tab:sensitivity}. For the sake of argument, we assign to each $f$/$g$ mixed mode multiplet a measurement error that is ten times that of the lowest-frequency multiplet of that degree, to simulate the difficulties we foresee in measuring them in the first place. All kernels shown are then constructed with values of the uncertainty trade-off parameter $\mu$ selected to yield a measurement uncertainty of $100$ nHz in the cyclic rotational frequency $\Omega / 2\pi$. For each kernel we compute its actual location as given by its centre of sensitivity,
\begin{equation}
    \left<r\right>_i = \int r \,K_{\text{eff},i} (r)\ \mathrm d r.\label{eq:centroid}
 \end{equation}
 The localisation uncertainty of each kernel is given by the spread around its centroid, which we estimate as
\begin{equation}
    \sigma_i \equiv \sqrt{\int r^2\,K_{\text{eff},i}(r)\ \mathrm{d}r -\left<r\right>_i^2}, \label{eq:spread}
\end{equation}
and use this measure for the error intervals and bars in Fig.\ref{fig:avg_kernels}.

In the top panel, we compare the locations and spreads of kernels constructed using only pure $p$ modes (orange), against those including $f/g$ mixed modes (blue). The latter bring no appreciable benefit outside a target radius of $0.2 \Rsun$, but we find that at about $r_0 = 0.07 \Rsun$, it is possible to construct a localisation kernel that minimises (in a local sense) both the discrepancy between the target and actual location of the kernel, as well as the spread of the kernel around its actual localisation. Fortuitously, this is also where the location of the centroid is closest to the centre. We examine this in more detail in the middle panel of \autoref{fig:avg_kernels}, where we compare mixed-mode and $p$ mode kernels targeted at this location (blue), and at $0.2 \Rsun$ (orange), which in turn is roughly where the centroids of pure $p$ mode kernels are closest to the centre. While the mixed and $p$ modes kernels exhibit minimal differences in shape at $0.2 \Rsun$, they differ substantially at $0.07 \Rsun$. At that target location, the pure $p$ mode kernel has a centre of sensitivity which actually lies substantially outwards of $r=0.2 \Rsun$. By contrast, the mixed-mode kernel has a centre of sensitivity which lies much closer to its target location.


In order to confirm that they do map the rotation profile of the rotation rate in two different radii, we simulated the inversion of the rotation profiles corresponding to the two scenarios of \citet{fuller2019} and \citet{eggenberger2019rotation} presented in Sect. \ref{sec:introduction}. The first has a flat rotation profile $\Omega(r)$ = cst in the whole radiative interior, the latter has an increasing rate towards the centre $r\to0$. In each case, we compute the integrals $\Omega_{f/g} \equiv \int \Omega(r) K_{f/g}(r)\;\mathrm{d}r$ and $\Omega_p \equiv \int \Omega(r)K_p(r)\;\mathrm{d}r$ {for the two averaging kernels we constructed using only $p$ modes (denoted $K_p(r)$) or $p$ and $f/g$ modes ( denoted $K_{f/g}$) respectively.} 
For both scenarios, we find the reasonable agreements $\Omega_{f/g} \simeq \Omega(r/\Rsun = 0.07)$ and $\Omega_p \simeq \Omega(r/\Rsun = 0.2)$, confirming that the two average kernels do provide independent measures of the rotation rate at these two radii. These results are shown in the bottom panel of Fig.~\ref{fig:avg_kernels}.

\section{Constraint with only one mode}

As mentioned in Sect. \ref{sec:mixed}, the detection of oscillation modes is more difficult at lower frequencies due to the signal of convection that becomes stronger. It is for this reason that we adopted adversarially high values for the notional measurement uncertainties for the splittings of these $f/g$ mixed modes multiplets. We should however also consider the possibility of early, partial detections of mixed $f$/$g$ modes, only at high frequencies. From \autoref{tab:sensitivity}, it appears that the modes with angular degrees $\ell$ > 8 cannot be used to probe the solar core as no significant coupling between $f$ and $g$ modes is found, and can thus be discarded. 
Taking that into account, the most probable detection with measurable sensitivity to rotation of the core is the mode with $\ell$ = 7 and $\nu$ = 305 $\mu$Hz. It is strongly $f$ dominated, which is favourable for high signal-to-noise ratio at the surface, as confirmed by the low $E_{\rm norm}$ value, while still having a non-zero mixing fraction and a sensitivity $s=0.06$. Although the $\ell = 7$ $f/g$ rotational splittings were assigned a measurement uncertainty of $\Delta\omega/2\pi = 56$ nHz in our MOLA exercise --- ten times that of currently measurable multiplets --- such core sensitivity would still permit this mode to estimate the core rotation if it were measurable at a precision of $25$ nHz (i.e. five times the current observational uncertainty). Under this scenario, we would obtain $\sigma_{\Omega_\mathrm{core}}/2\pi = 25/0.06$ nHz $ = 420$ nHz.
While such a single-multiplet detection would provide only a single value on the solar core rotation rate, and with a notable uncertainty, it would still suffice to discriminate between the predictions of \citet{fuller2019} and \citet{eggenberger2019rotation}, as this uncertainty is less than their disagreement. 
Indeed, the almost solid-body rotation predicted by \citet{fuller2019} predicts $\Omega_\mathrm{core}/2\pi = 440$nHz, and the radial differential rotation predicted by \citet{eggenberger2019rotation} implies rotation rates reaching more than $1000$nHz. Therefore, an average measure of the rotation rate profile in these layers with uncertainty of $420$ nHz would suffice to distinguish between these two scenarios.

\section{Discussion}
Our results demonstrate that mixed $f/g$ modes can provide precise constraints on the rotation rate of the solar core. First, we show that in addition to the mixed $f$/$g$ mode at an angular degree $\ell $= 4 predicted by \citet{LeSaux2025}, the oscillation spectrum of the Sun might contain up to 12 mixed $f$/$g$ modes for angular degrees between $\ell$ = 3 and 8. Modes with higher degrees, $\ell$ > 8, are unlikely to couple in the Sun and/or are not sensitive to the dynamics of the core. Then, we combine the probing power of the lower degree modes, which are at the same time sensitive to the core and should have significant amplitude at the surface, to show that it is possible to get two independent measurements of the rotation rate of the Sun's core at two different radii. To do so, we build an {averaging kernel}, which is a function estimating how sensitive modes are at a given radius, including these mixed modes to the currently observed $p$ modes.
As illustrated on Fig. \ref{fig:avg_kernels}, with this inclusion we managed to obtain a kernels peaking well within the solar core. Detection of these few mixed modes would then be enough to discriminate between two competing angular momentum transport scenarios proposed by \citet{fuller2019} and \citet{eggenberger2019rotation}, which predict a solid-body rotation for the former and a fast rotating core for the latter.

We have also shown that the splitting of the highest-frequency of these mixed modes (at $\ell = 7$ and $\nu = 305\,\mu$Hz) is sufficient to discriminate between these two scenarios as well. It would however be insufficient to actually invert the rotation profile in the core.
While the frequency resolution required for such a detection is already accessible with current data from GOLF or BiSON, the main obstacle comes for the convective noise dominating the signal at low frequencies.
In \citet{LeSaux2025}, we estimated that the order of magnitude for the surface amplitude of mixed $f$/$g$ modes is the same as for $p$ modes in a similar frequency range ($\sim$ 300 $\mu$Hz). As argued in \citet{Davies2014}, the perspective of detecting such low-frequency modes remains quite low. With just continuous observations, reaching the detection threshold at these frequencies might require at least 30 or 40 more years of data \citep[see Fig. 6 in][]{Davies2014}. However, a more optimistic perspective might be to observe the Sun at multiple wavelengths, which allow probing the solar atmosphere at different heights. In particular, $p$ modes have been detected and characterised by \citet{Howe2011} with the Atmospheric Imaging Assembly (AIA) onboard the Solar Dynamics Observatory (SDO). AIA observes in near-UV wavelength and probe directly the solar chromosphere where the noise from granulation seems weaker at low frequencies. Such a method mights offer interesting opportunities to detect mixed $f$/$g$ modes in the near future.


\begin{acknowledgements}
      AL acknowledges funding from ERC StG project Calcifer No 101165631. ALS acknowledge support from the European Research Council (ERC) under the Horizon Europe programme (Synergy Grant agreement 101071505: 4D-STAR). JMJO acknowledges support from the Australian Research Council (FL220100117 and FT200100871), and from NASA through the NASA Hubble Fellowship HST-HF2-51517.001, awarded by STScI. STScI is operated by the Association of Universities for Research in Astronomy, Incorporated, under NASA contract NAS5-26555. While partially funded by the European Union, views and opinions expressed are however those of the authors only and do not necessarily reflect those of the European Union or the European Research Council. Neither the European Union nor the granting authority can be held responsible for them. RAG acknowledges the support from the GOLF and PLATO Centre National D'{\'{E}}tudes Spatiales grants.
\end{acknowledgements}

\bibliographystyle{aa}
\typeout{}
\bibliography{bib}{}

\appendix

\section{$\varphi/\gamma$ Decomposition}
\label{app:decomp}

In other asteroseismic contexts in which mixed modes emerge, e.g. for mixed modes in red giants, an asymptotic JWKB description is often used. However, $f$ modes (being definitionally $n=0$ modes) are poorly described by the JWKB approximation both qualitatively and quantitatively. As an alternative to this, \cite{Ong2020} describe a procedure for isolating $p/g$ mixed modes into linear combinations of notional pure $p$ and pure $g$ modes, obtained by suppressing terms of the wave operator when solving for normal modes of a boundary eigenvalue problem. The decoupled solutions so obtained are referred to as $\pi$ and $\gamma$ modes, to distinguish them from pure $p$ and $g$ modes obtained from the unmodified wave operator. This nomenclature is inherited from \cite{Aizenman1977}, wherein it was used to describe a different isolation scheme (with different terms suppressed in a manner more appropriate for studying high-mass stars). For the isolation scheme of \cite{Ong2020}, the mixed-mode frequencies and eigenfunctions can be found by solving a Generalised Hermitian Eigenvalue Problem of the form
\begin{equation}
    \left(
    \begin{bmatrix}
    \mathbf{L}_{\pi\pi} + \mathbf{R}_{\pi\pi} & \mathbf{L}_{\pi\gamma} + \mathbf{R}_{\pi\gamma} \\
    \mathbf{L}_{\pi\gamma}^\dagger + \mathbf{R}_{\pi\gamma}^\dagger & \mathbf{L}_{\gamma\gamma} + \mathbf{R}_{\gamma\gamma}
    \end{bmatrix} + \omega^2
    \begin{bmatrix}
    \mathbb{I} & \mathbf{D}_{\pi\gamma} \\
    \mathbf{D}_{\pi\gamma}^\dagger & \mathbb{I}
    \end{bmatrix}
    \right)
    \begin{bmatrix}
    \mathbf{c}_\pi \\ \mathbf{c}_\gamma
    \end{bmatrix} = 0.
    \label{eq:eigenproblem}
\end{equation}
We refer the reader to \cite{Ong2020} for a complete derivation, and for explicit expressions for these matrix elements in terms of overlap integrals between the mode eigenfunctions and various operators. For our purposes, the following properties are relevant:
\begin{itemize}
    \item The on-block-diagonal matrices $\mathbf{L}_{\pi\pi}$ and $\mathbf{L}_{\gamma\gamma}$ are separately diagonal, and their entries $-\omega_i^2$ are specified by the frequencies of the $\pi$ and $\gamma$ modes obtained directly from the pulsation calculations involving modified wave operators.
    \item The off-block-diagonal matrices $\mathbf{L}_{\pi\gamma}$, $\mathbf{R}_{\pi\gamma}$, and $\mathbf{D}_{\pi\gamma}$ are specified by overlap integrals between one $\pi$ and one $\gamma$ mode each, and collectively specify the strength of the coupling between the two families of modes. \cite{Ong2023} in turn derive expressions translating between these overlap integrals, and both the dimensionless coupling strengths more commonly used in the JWKB approximation, as well as a characteristic resonance width $\Gamma$, \autoref{eq:resonance_width} (so that two modes are considered close to resonance if their separation is less than $\Gamma$).
    \item The frequencies of the isolated pure $p$ and $g$ modes that would be associated with the unmodified wave operator can be well approximated by the diagonal elements of $\mathbf{L}_{\pi\pi} + \mathbf{R}_{\pi\pi}$ and $\mathbf{L}_{\gamma\gamma} + \mathbf{R}_{\gamma\gamma}$, respectively, to first order in perturbation theory.
    \item  The eigenvalues $\omega^2$ of this problem yield the mixed-mode frequencies, while the corresponding eigenvectors $\mathbf{c}$ specify coefficients permitting each mixed-mode eigenfunction to be expressed as a linear combination of these $\pi$ and $\gamma$ modes,
    \begin{equation}
    \boldsymbol{\xi}_\text{mixed} = \sum_i \boldsymbol{\xi}_{\pi, i} c_{\pi, i} + \sum_i \boldsymbol{\xi}_{\gamma, i} c_{\gamma, i}.\label{eq:combinations}
\end{equation}
\end{itemize}

In addition to $p$-like modes, \cite{Aizenman1977} found that their $\pi$-mode operator also yielded an $f$-like mode, which they referred to as the $\varphi$ mode, as its fundamental mode. Within the frequency range being considered in this work, we find that the isolation scheme of \cite{Ong2020} also gives rise to a $\varphi$ mode. In the main text of the paper, we report an effective pure $f$ mode frequency computed by evaluating \modif{the lowest diagonal element of} $L_{\pi\pi} + R_{\pi\pi}$ directly using the same first-order expressions as derived in \cite{Ong2020}. As can be seen in \autoref{fig:delta_pi}, these correspond quite well, qualitatively, to the locations of the dips that are obtained when taking pairwise consecutive period differences. Similarly, we evaluate the off-diagonal elements of \autoref{eq:eigenproblem} in order to compute a dimensionless coupling strength $q$ and frequency resonance width $\Gamma$ (\autoref{fig:resonance-widths}), using expressions derived in \cite{Ong2023}, which also well-describe the widths of the dips seen in \autoref{fig:delta_pi}. This suggests that these $\varphi$ modes do indeed well describe the pure $f$ modes of our solar model, in entirely analogous fashion to the $\pi$-modes for $p$ modes in red giant models. We rely on this ansatz to justify our computation of other mixed-mode quantities (e.g. mixing fractions $\zeta$) also using nonasymptotic expressions for them from \cite{Ong2020}.

To further validate this ansatz, we seek to show that the $f$/$g$ mixed-mode eigenfunctions can also be written as linear combinations of $\varphi$ and $\gamma$ modes. We do this by numerical demonstration, specifically for several mixed modes in the vicinity of the $f$-dominated pair that we describe at $\ell = 4$ in the main text. In order to better display the structure of these eigenfunctions both close to the core and close to the surface, we follow \cite{Tassoul1980} in using an asymptotic radial phase coordinate of the form
\begin{equation}
\begin{aligned}
    \theta(r; \omega) &= \int_0^r \sqrt{{\ell(\ell + 1)N^2 \over r'^2 \omega^2} + {\omega^2 \over c_s^2}}\ \mathrm d r'; \\
    y & = \theta(r; \omega) / \theta(R, \omega) \in [0, 1].
    \label{eq:tassoulcoord}
\end{aligned}
\end{equation}
The frequency dependence of this quantity controls the fraction of the coordinate which is sensitive to the core vs. the near-surface layers. For display purposes, we choose a value of $\omega$ so that, roughly speaking, the inner half is sensitive to the radiative interior and the outer half is sensitive to the near-surface layers. Also for display purposes, we rescale the eigenfunctions by the integral measure. We note that scaling both sides of \autoref{eq:combinations}, and changing the coordinate in which it is displayed, does not modify the property that each mixed mode can be approximated as a linear combination of decoupled basis modes.

\begin{figure*}[htbp]
    \centering
    \includegraphics[width=\textwidth]{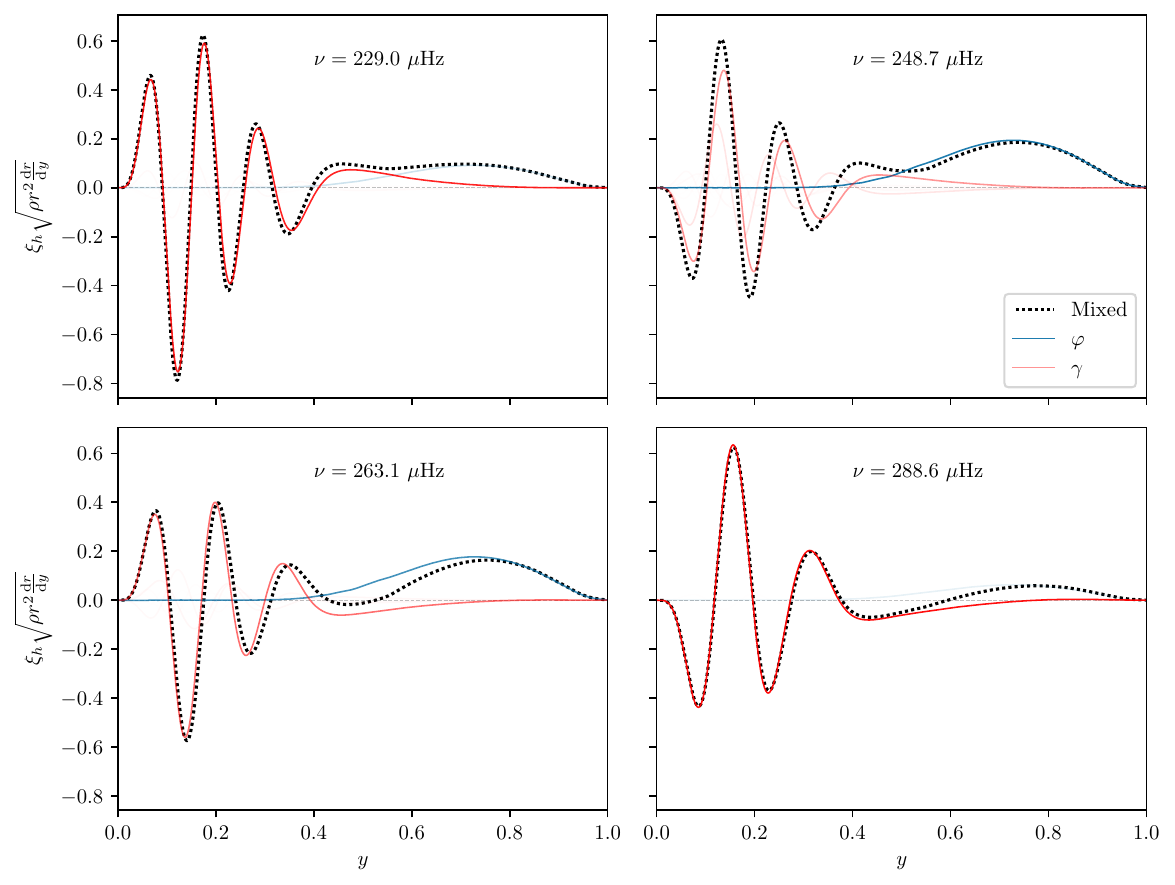}
    \caption{$\varphi/\gamma$ decomposition of mixed-mode eigenfunctions for several of the $\ell = 4$ $f$/$g$ mixed modes of the solar model considered in this work. All eigenfunctions shown are scaled by the change of coordinates described in \autoref{eq:tassoulcoord}. The scaled mixed-mode eigenfunctions are shown with the dotted black curves, $\gamma$ mode eigenfunctions with solid red curves, and $\varphi$ mode eigenfunctions with solid blue curves. The latter two are assigned opacities based on the squares of their coefficients $c_i$ in the description of each mixed mode as linear combinations $\xi_\text{mixed} = \sum c_i \xi_i$ of isolated $\varphi$ and $\gamma$ modes.}
    \label{fig:decomp}
\end{figure*}

We show both the mixed-mode and constituent $\varphi/\gamma$-mode eigenfunctions in \autoref{fig:decomp}. Since we are dealing primarily with $g$ modes, we choose to show the horizontal components of the eigenfunctions. The mixed-mode eigenfunctions are traced out with dotted black curves, while the pure $\varphi$ and $\gamma$-mode eigenfunctions are shown with solid thin blue and red curves, respectively, each assigned opacity given by the square of their coefficient $c_i^2$ in the linear combination constituting each mixed mode. The qualitative agreement is excellent, further cementing our ansatz.

\section{Rotational Averaging Kernels}
\label{app:coefs}


We employ a modified version of the method of optimal localised averages (OLA), in which we choose coefficients $c_i$ of the inversion kernels which optimise a penalty function $\mathcal{C}$ {which we specify below}, subject to additional constraints and further penalisation by observational uncertainties. {In this appendix, we use normalised kernels defined as $\mathcal{K}_i(r) \equiv \frac{1}{\beta_i}K_{\ell_i,m_i,n_i}(r)$ for the kernel of a given mode (${\ell_i,m_i,n_i}$) with $\beta_i \equiv \int K_{\ell_i,m_i,n_i}(r)\mathrm{d}r$. Therefore, $\mathcal{K}_i$ is of unit integral.} Generically, the constrained problem involves optimising an auxiliary penalty function
\begin{equation}
\Lambda(\{c_i\}; \alpha, \gamma) = \mathcal{C} + \mu\left(\sum_{i,j}c_i c_j \Sigma_{ij}\right) + \alpha \left(\sum_i c_i\beta_i - 1\right) + \gamma \left(\sum_i c_i\eta_i\right).   
\end{equation}
This is a Lagrange multiplier problem, with $\mu$ being a tradeoff parameter between tightness of localisation and the uncertainty of the inversion result with $\mathbf{\Sigma}$ being the error-covariance matrix, and the Lagrange multiplier $\alpha$ enforcing a unit integral in the combined kernel (as is usual practice). In our case, we also have a second Lagrange multiplier $\gamma$ enforcing a desired level of contributions from mixing with $f$ or $p$ modes, as in \cite{one_inversion_2024} for mixing between $p$ and $g$ modes. {$\eta_i$ is a flexible notation to specify what one wants to minimize. Aiming for the core, our averaging kernels need the $g$ part of mixed modes, such that the factors to be minimized are $\eta_i = \beta_{i}(1 - \zeta_i)$, i.e \modif{ensuring} high mixing fractions}. Optimal solutions then satisfy
\begin{equation}
\begin{aligned}
0 = {\partial \Lambda \over \partial c_i} &= {\partial \mathcal C \over \partial c_i} + 2\mu \sum_j \Sigma_{ij} c_j + \alpha \beta_i + \gamma \eta_i, \\
0 = {\partial \Lambda \over \partial \alpha} &= \sum_i c_i\beta_i - 1,\\
0 = {\partial \Lambda \over \partial \gamma} &= \sum_i c_i\eta_i.
\end{aligned}    
\end{equation}

For SOLA \citep{pijpers_sola_1994}, the penalty function is specified by the $L^2$ norm:
\begin{equation}
    \mathcal{C} = \int\left(\sum_i c_i \beta_i \mathcal{K}_i - \mathcal{T}\right)^2\mathrm dx \text{ and } {\partial \mathcal C \over \partial c_i} = 2 \int \left(\sum_j c_j \beta_j \mathcal{K}_j - \mathcal{T}\right)\beta_i \mathcal{K}_i \mathrm d x,
\end{equation}
allowing us to write the problem in block matrix form as
\begin{equation}
\begin{bmatrix}
2(\mathbf{A} + \mu \mathbf\Sigma) & \mathbf{\beta} & \mathbf{\eta} \\
\mathbf{\beta}^T & 0 & 0 \\
\mathbf{\eta}^T & 0 & 0
\end{bmatrix}
\begin{bmatrix}
\mathbf{c} \\ \alpha \\ \gamma
\end{bmatrix}
= 
\begin{bmatrix}
\mathbf{B} \\ 1 \\ 0
\end{bmatrix},\label{eq:olainverse}
\end{equation}
where $A_{ij} = \beta_i \beta_j \int \mathcal{K}_i \mathcal{K}_j\mathrm d x$, $B_i = \beta_i \int \mathcal{K}_i \mathcal{T} \mathrm d x$, and $\mathcal{T}$ is the target kernel. The vector of inversion coefficients, $\mathbf{c}$, is then to be found from this linear inverse problem. The estimated uncertainty in the inverted rotation rate is given by $\sigma_i^2 = \mathbf{c}^T \mathbf{\Sigma} \mathbf{c}$.

For MOLA \citep{backus1968}, we have instead that
$$
\mathcal{C} = \int \mathcal{W}(x) \left(\sum_i c_i \beta_i \mathcal{K}_i\right)^2\ \mathrm d x \text{ and } {\partial \mathcal{C} \over \partial c_i} = 2\int \mathcal{W} \left(\sum_j c_j \beta_j \mathcal{K}_j\right)\beta_j \mathcal{K}_j\ \mathrm d x
$$
where $\mathcal W$ is a weight function (chosen to be close to 0 near the target location and large far away from it). A common choice is $\mathcal{W}(x) = (x - x_0)^2$. The coefficients $\{c_i\}$ are then found from solving the same linear problem as \autoref{eq:olainverse} with
\begin{equation}
    A_{ij} = \int \mathcal{W}\beta_i \mathcal{K}_i \beta_j \mathcal{K}_j\ \mathrm d x \text{, and  } \mathbf{B} = 0.
\end{equation}

In the main text, we present analysis using only localisation kernels constructed using MOLA, in order to simplify the analysis. We do this because the SOLA procedure also requires us to calibrate an additional free parameter specifying the widths of target kernels. However, we obtain qualitatively similar results when using SOLA, although the details (e.g. location of the optimally-located kernel) do depend on the width parameter chosen.

\end{document}